\documentstyle[amsfonts,epsfig,12pt]{article}

\def\mypagenumber{1}

\def\myend{\end{document}}

\normalsize

\newcounter{sxn}

\newcounter{axn}

\date{}

\newdimen\mybaselineskip
\def\fft#1#2{{#1 \over #2}}

\newcommand{\beeq}{\begin{equation}}
\newcommand{\eneq}{\end{equation}}
\newcommand{\be}{\begin{eqnarray}}
\newcommand{\ee}{\end{eqnarray}}
\newcommand{\bpic}{\begin{picture}}
\newcommand{\epic}{\end{picture}}

\def\la{\raise.16ex\hbox{$\langle$} \, }
\def\ra{\, \raise.16ex\hbox{$\rangle$} }

\def\psibar{ \psi \kern-.65em\raise.6em\hbox{$-$} }
\def\mbar{ m \kern-.78em\raise.4em\hbox{$-$}\lower.4em\hbox{} }

\def\L{ {\Lambda} }
\def\a{ {\alpha} }
\def\m{ {\mu} }

\def\b{ {\beta} }
\def\L{ {\Lambda} }
\def\ep{\epsilon}

\def\n@space{\nulldelimiterspace=0pt \mathsurround=0pt }
\def\huge#1{{\hbox{$\left#1\vbox to 20.5pt{}\right.\n@space$}}}

\def\myskip{\noalign{\kern 8pt}}
\def\myeqspace{\noalign{\kern 10pt}}

\def\boxit#1{$\vcenter{\hrule\hbox{\vrule\kern3pt
    \vbox{\kern3pt\hbox{#1}\kern3pt}\kern3pt\vrule}\hrule}$}
\def\bigbox#1{$\vcenter{\hrule\hbox{\vrule\kern5pt
     \vbox{\kern5pt\hbox{#1}\kern5pt}\kern5pt\vrule}\hrule}$}

\def\ignore#1{{}}

\renewcommand{\baselinestretch}{1.0}

\begin{document}
\bibliographystyle{unsrt}
\footskip .5cm

\thispagestyle{empty}
\setcounter{page}{\mypagenumber}


\begin{flushright}{
BRX-TH-566\\}

\end{flushright}

\vspace{2.5cm}
\begin{center}
{\LARGE \bf {Conserved Charges of Higher $\mbox{\boldmath $D$}$ \\\vspace{.2in} Kerr--AdS Spacetimes}}\\
\vskip 1 cm {\large{S. Deser,$^a$ Inanc Kanik$^b$ and  Bayram
Tekin$^b$}} \\
\vspace{.5cm} {\it $^a$Department of  Physics,
Brandeis University, Waltham, MA 02454,
USA}\\{\tt deser@brandeis.edu}\\
\vskip 0.3 cm
 {\it{$^b$Physics Department, Middle East Technical University, 06531
Ankara, Turkey}\\{\tt btekin@metu.edu.tr}} \\

\renewcommand\baselinestretch{2.0}

\begin{abstract}
We compute the energy and angular momenta of recent
$D$-dimensional Kerr-AdS solutions to cosmological Einstein
gravity, as well as of the BTZ metric, using our invariant charge
definitions.
\end{abstract}

\end{center}

\vspace*{1.5cm}

\newpage

\normalsize

 \baselineskip=22pt plus 1pt minus 1pt
\parindent=25pt
\vskip 2 cm

\section{Introduction}

Rotating solutions of cosmological Einstein gravity in $D$
dimensions, $R_{\mu\nu} = (D-1)\,\Lambda\, g_{\mu\nu}$, have been
constructed recently \cite{gibbons1, gibbons2}, extending earlier
$\L =0$ solutions of \cite{myers}, themselves generalizations of
the well-known $D$=4 metrics of \cite{carter} and \cite{kerr}, and
of \cite{hawking} in $D$=5. These geometries provide a useful
application of our recent generalized ``conserved charge"
definitions, which are also extensions -- of the original ADM
\cite{adm}, and AD \cite{ad} charges -- to cover wider classes of
actions \cite{dt1,dt2}: We will compute the energy and angular
momenta of these new solutions, as well as of the $D=3$ BTZ metric
as calculated within topologically massive gravity.

Gravity theories have been historically endowed with a variety of
seemingly different charge definitions, with different degrees of
applicability and coordinate invariance. This topic has also seen
much very recent activity, for example \cite{petrov}. A summary and
comparison of some of them is given in \cite{pope} which also
includes a computation of the charges for Kerr-AdS black holes,
using thermodynamic arguments; see also \cite{deruelle,katz}. Our
results will agree with those, but we emphasize that in a general
context, certain coincidences between charge definitions are
suspect: For example, the frequently invoked ``Komar" charges, are
in general not applicable, being highly gauge-dependent
\cite{misner}.

\section{Mass and Angular Momenta of Kerr-AdS}

Let us briefly recapitulate the formulations of \cite{ad, dt1}. The field
equations of any metric model coupled to a (necessarily covariantly conserved) matter source
$\tau_{\mu \nu}$ are
 \be
  {\delta I \over \delta  g_{\mu \nu}} \equiv  \Phi_{\mu
\nu}(g, R, \nabla R, ..) = \kappa \tau_{\mu \nu}, \label{generic}
 \ee
where $\Phi_{\mu\nu}$ is an identically conserved tensor that can
depend on curvatures and their derivatives. Decompose the metric
into the sum of a background ``vacuum", $\bar{g}_{\mu \nu}$ (which
solves (\ref{generic}) for $\tau_{\mu \nu} = 0$), plus a deviation
$h_{\mu \nu}$, not necessarily small, that vanishes sufficiently
rapidly far from the matter source: $g_{\mu \nu}= \bar{g}_{\mu
\nu}+ h_{\mu \nu}$. The field equations can be divided into a part
linear in $h_{\mu \nu}$ plus a non-linear remainder, which (with
$\tau_{\mu\nu}$) constitutes the total source $T_{\mu \nu}$. If
the background $\bar{g}_{\mu\nu}$ admits Killing vectors
$\bar{\xi}_{\mu}$, obeying ${\bar{\nabla}}_\mu \bar{\xi}_\nu +
{\bar{\nabla}}_\nu \bar{\xi}_\mu = 0 $, then, up to normalization
factors (which we shall fix later), the conserved Killing charges
are
 \be
Q^\mu(\bar{\xi}) = \int_{\cal{M}} d^{D-1} x \sqrt{-\bar{g}} T^{\mu
\nu}\bar{\xi}_\nu =  \int_{\Sigma} dS_i {\cal{F}}^{\mu i}.
\label{charge}
 \ee
Here  $\Sigma$ is a $D-2$ dimensional space-like asymptotic hypersurface
of the space $\cal{M}$ and ${\cal{F}}^{\mu i}$ is an anti-symmetric tensor,
whose explicit form is model-dependent. For Einstein's theory with a cosmological constant,
 \be
 Q^{\mu} = {1\over 4
\Omega_{D-2} G_D}\int_{\Sigma} &dS_i&\Big \{ \bar{\xi}_\nu
\bar{\nabla}^{\mu}h^{i \nu} -\bar{\xi}_\nu
\bar{\nabla}^{i}h^{\mu\nu} +\bar{\xi}^\mu \bar{\nabla}^i h
-\bar{\xi}^i \bar{\nabla}^\mu h \nonumber \\
&& + h^{\mu \nu}\bar{\nabla}^i \bar{\xi}_\nu - h^{i
\nu}\bar{\nabla}^\mu \bar{\xi}_\nu + \bar{\xi}^i
\bar{\nabla}_{\nu}h^{\mu \nu} -\bar{\xi}^\mu
\bar{\nabla}_{\nu}h^{i \nu} + h\bar{\nabla}^\mu \bar{\xi}^i \Big
\} \; , \label{ad}
 \ee
 where $i$ takes values in $1, 2,... D-2$ and
the charge is normalized as shown, by dividing with the $D$-dimensional
Newton's constant and the solid angle. These charges are
background gauge invariant under the diffeomorphisms $
\delta_\zeta h_{\mu \nu} = \bar{\nabla}_\mu \zeta_\nu +
\bar{\nabla}_\nu \zeta_\mu $:   $\delta_\zeta Q^\mu = 0$.

Let us now calculate the conserved charges of the metrics
\cite{gibbons1} for $D > 3$. [We shall treat the special $D= 3$ case
at the end]. They have the Kerr-Schild form
\cite{kerrschild1,gursey}
 \be
 ds^2 = d\bar s^2 +
\fft{2M}{U}\, (k_\mu\, dx^\mu)^2 \; ,\label{kschild}
 \ee
 in terms of the de Sitter metric
  \be d\bar s^2 &=& -W\,(1-\Lambda \,
r^2)\, dt^2 + F\, dr^2 + \sum_{i=1}^{N+\ep} \fft{r^2 +
a_i^2}{1+\Lambda\, a_i^2} \,\, d\mu_i^2 + \sum_{i=1}^N \fft{r^2 +
a_i^2}{1+\Lambda\, a_i^2}
\, \, \mu_i^2\, d\phi_i^2 \nonumber \\
&&+ \fft{\Lambda}{W\, (1-\Lambda\, r^2)}\, \Big( \sum_{i=1}^{N+\ep}
\fft{(r^2 + a_i^2)\, \mu_i\, d\mu_i}{1+\Lambda\, a_i^2}
   \Big)^2 \; . \label{ds}
\ee
 Here  $\epsilon  = 0/1$ for odd/even, dimensions and  $ D = 2N + 1
 + \epsilon$. The null 1-form reads
 \be
 k_\mu\, dx^\mu &=& F\, dr + W\, dt - \sum_{i=1}^N \fft{a_i\,
\mu_i^2}{1+\Lambda\, a_i^2} \,  d\phi_i \; , \ee
 with
  \be
 U &\equiv& r^{\ep}\, \sum_{i=1}^{N+\ep} \fft{\mu_i^2}{r^2 +
a_i^2}\, \prod_{j=1}^N (r^2 + a_j^2), \,\, \,   W \equiv
\sum_{i=1}^{N+\ep} \fft{\mu_i^2}{1+\Lambda\, a_i^2}, \,\,\, F\equiv
\fft{1}{1-\Lambda\, r^2}\, \,
  \sum_{i=1}^{N+\ep} \fft{r^2 \, \mu_i^2}{r^2+a_i^2} \; .
\ee
 To find the energy and angular momenta corresponding to
(\ref{kschild}), we must compute the charges $Q^0$ for the
corresponding Killing vectors: for the energy we shall take $
\bar{\xi}^\mu = (-1, \vec{0})$  and each angular momentum has the
appropriate unit entry $(0,\ldots 1_i\ldots0)$.  Then
 \be
 Q^0 &=& {1\over 4 \Omega_{D-2}
G_D}\int_{\Sigma} dS_r \Big\{
 g_{00} \bar{\nabla}^{0}h^{r 0} +g_{00} \bar{\nabla}^{r}h^{00} +
h^{0 \nu}\bar{\nabla}^r \bar{\xi}_\nu - h^{r \nu}\bar{\nabla}^0
\bar{\xi}_\nu  + \bar{\nabla}_{\nu}h^{r \nu} \Big \} \; .
 \ee
Using the energy Killing vector, we obtain\footnote{We are assuming
that the background spacetime is AdS rather than dS, whose
cosmological horizon causes complications. Some of these issues were
addressed in \cite{ad,dt1}. For details of acceptable asymptotic
falloff to (A)dS in various dimensions, we refer to \cite{hen}. }
 \be
  E_D &=& {1\over 4 \Omega_{D-2}
G_D}\int_{\Sigma} dS_r \Big \{ g_{00}g^{rr}\partial_r
h^{00}+\frac{1}{2}h^{00}g^{rr}\partial_r g_{00}-
\frac{m}{U}g^{00}\partial_r g_{00}+2m\partial_r U^{-1}\nonumber\\& &
\hskip 3 cm +\frac{2m}{U}g^{rr}\partial_r g_{rr} -\frac{m}{U}g^{rr}
k^ik^j\partial_r g_{ij}+\frac{m}{U} g^{ij}\partial_r g_{ij}\Big \}.
\label{ed}
 \ee
To compute $E_D$, one needs the large $r$ behavior of the integrand
$I$ of (\ref{ed}); since
 \be
  g_{00}\rightarrow W \Lambda r^2,
\hskip 1 cm F \rightarrow \frac{-1}{\Lambda r^2}, \hskip 1 cm
U\rightarrow r^{D-3}, \hskip 1 cm  k^\phi&\rightarrow {a_\phi \over
r^2} \; , \label{larger}
 \ee
then
  \be
  I=\frac{2m}{r^{D-2}}[(D-1)W-1] \;.
\label{integrand}
 \ee
For completeness, let us also note how the determinant is
calculated,
  \be
{\mbox{det}}g=-W(1-\Lambda r^2) F\prod_{i=1}^{N}
\frac{(r^2+a_i^2)\mu_i^2}{1+\Lambda a_i^2}\,\mbox {det}M \; .
\label{fulldet}
 \ee
Here $M$ is the matrix representing the coefficients of the form
$d\mu_i d\mu_j$ in the
 metric, which can be expressed as (no repeated index summation),
\be
 M_{ij}=A_i \delta_{ij} + B_i B_j + C_i C_j  \hskip 1 cm
 \ee
where
 \be
A_i&=&\frac{(r^2+a_i^2)}{1+\Lambda a_i^2}, \hskip 2 cm B_i=
\sqrt{\frac{(r^2+a_{N+\epsilon}^2)}{1+\Lambda a_{N+\epsilon}^2}}
\frac{\mu_i}{\mu_n} \nonumber \\
C_i&=& \sqrt{ \frac{\Lambda}{W(1-\Lambda r^2)}}
(\frac{(r^2+a_i^2)}{1+\Lambda
a_i^2}-\frac{(r^2+a_{N+\epsilon}^2)}{1+\Lambda a_{N+\epsilon}^2})
\mu_i \; .    \label{funnybits}
 \ee
Then we have
 \be
{\mbox{det}}M=\prod_{i=1}^{N+\epsilon-1}A_i
\sum_{i=1}^{N+\epsilon-1}\Big \{ \frac{B_i^2}{A_i}+
\frac{C_i^2}{A_i}+\sum_{j\neq i}^{N+\epsilon-1}
  \frac{B_i^2 C_i^2}{A_i A_j}-\sum_{j\neq i}^{N+\epsilon-1}
  \frac{B_i B_j C_j C_i}{A_i A_j} \Big\} \; .
\ee
 Inserting (\ref{funnybits}) in the above equation, one gets
  \be
{\mbox{det}}M=\frac{1}{W \mu_{N+\epsilon}^2}
\prod_{i=1}^{N}\frac{1}{1+\Lambda a_i^2} \; . \label{determinant}
 \ee
Using equations (\ref{determinant},\ref{fulldet},\ref{integrand})
the energy of the $D$ dimensional rotating black hole becomes
 \be
E_{D}= {m \over \Xi} \sum_{i=1}^{{D-1-\epsilon\over 2}} \left \{{1
\over \Xi_i}- (1-\epsilon)({1\over 2}) \right \}.\label{finalen} \ee
where \be \Xi \equiv \prod_{i=1}^{ {D-1-\epsilon\over 2}} (1+
\Lambda a_i^2), \, \, \,\,\, \Xi_i \equiv {1+ \Lambda a_i^2}.
 \ee
This expression reduces to the standard limits $ a_i \rightarrow 0$
and $\Lambda \rightarrow 0$, and agrees (up to a constant factor)
with those of \cite{pope,deruelle}.

The computation of angular momenta follows along similar lines.
Consider a given, say that $i^{\mbox{th}}$ (which we call the
$\phi$) component, {\it i.e.}, the Killing vector
$\xi^{\mu}_{(i)}=(0,...,0,1_i,0,..)$. Then the corresponding
Killing charge becomes
 \be Q^{0} &=& {1\over 4 \Omega_{D-2}
G_D}\int_{\Sigma} dS_r \Big \{ g_{\phi\phi} \bar{\nabla}^{0}h^{r
\phi} -g_{\phi\phi} \bar{\nabla}^{r}h^{0\phi} + h^{0
\nu}\bar{\nabla}^r \bar{\xi}_\nu - h^{r \nu}\bar{\nabla}^0
\bar{\xi}_\nu   \Big \}\nonumber\\
&=& {1\over 4 \Omega_{D-2} G_D}\int_{\Sigma} dS_r \Big \{
-g_{\phi\phi}g^{rr}g^{00}\partial_r h_0^{\;\phi} \Big \} \; .
 \ee
Once again the integrand can be calculated to be
 \be
  I = {(D-1)2m a_i \mu_i^2\over r^{D-2}( 1+ \Lambda a_i^2)}.
 \ee
Putting the pieces together, the angular momentum is
 \be
 J_i =  { m a_i\over \Xi \Xi_i }.
\label{ang}
 \ee
This expression again agrees with \cite{pope,deruelle}. Note that,
unlike in the energy expression, $\epsilon$ does not appear here
since even dimensional spaces have as many independent 2-planes as
the odd dimensional spaces with one lower dimension.\footnote{For
even dimensions, there is a nice relation between the energy and the
angular momentum $ E= \sum_i {J_i \over a_i}$.}

Having computed the desired conserved charges (17,21) for Kerr-AdS
spacetimes in $D >3$, let us briefly turn our attention to the $D =
3$ BTZ black hole \cite{btz}.  This solution has long been studied
but we recompute the charges with our method for the sake of
completeness. The BTZ black hole differs from its higher dimensional
counterparts in one very important aspect: for it, AdS is not the
correct-vacuum-background \cite{btz}. The full metric is
 \be
 ds^2 =
(M - \Lambda r^2)dt^2 + {dr^2\over {-M +\Lambda r^2 + {a^2\over {4
r^2}}}} -a dt\, d\phi + r^2 d\phi^2 \;,
 \ee
The background metric corresponds to  $M =0$ and AdS corresponds to
$M = -1$.  Only AdS with $ J = 0$ is allowed for $M<0$: the others
have naked singularities. So we consider $M >0$ and compute the
charges following our calculations above (about the $ M= 0$
background.) We get the usual answers
 \be
 E = M  \; , \hskip 2 cm J = a\; .
 \ee
BTZ black holes also solve the more general topologically massive
gravity equations, where the Einstein term is augmented by the
Cotton tensor \cite{djt},
 \be
 G_{\mu\nu} \, +\Lambda g_{\mu \nu} + {1\over \mu } \:
 C_{\mu \nu}= \kappa \tau_{\mu \nu} \; .
 \ee
Conserved charges in this model were obtained in \cite{dest}, in
terms of those of the Einstein model $Q^\m_E$,
 \be Q^\mu
 (\bar{\xi}) &=& Q^\mu_E (\bar{\xi})  + {1\over 2\mu}\oint
d S_i  \left \{ \epsilon^{\mu i \beta} {\cal{G}}^L\,_{\nu
\beta}\bar{\xi}^\nu +\epsilon^{\nu i}\,_\beta {\cal{G}}_L\,^{\mu
\beta}\bar{\xi}_\nu +\epsilon^{\mu \nu \beta} {\cal{G}}^{L\,
i}\,_\beta \bar{\xi}_\nu \right \}\\ \nonumber &&+{1\over 2\mu}\:
Q^\mu_E (\epsilon \bar{\nabla}\bar{\xi})\;, \label{chargeF}
 \ee
where $Q^\mu_E (\epsilon \bar{\nabla}\bar{\xi})$ is the Einstein
form but $\bar{\xi}$ is replaced with its curl. Once the
contributions of the Cotton parts are computed the mass and the
angular momentum of the BTZ black hole reads: \be E = M - {\Lambda
a \over \mu}, \hskip 1 cm J = a - {M\over \mu}, \ee a shift in
values that may be compared with those for gravitational anyons
\cite{deser}, (linearized) solutions of TMG but not of pure D=3
Einstein.

\section{Mass and Angular Momenta in Higher Curvature Models}

We turn now to a slightly more formal exercise, which is to
indicate the stability of our generic charge definition framework
as it applies to a wider range of models, specifically higher
derivative gravities. While Kerr-like solutions to $R+R^2$ gravity
models have yet to be discovered, it is not unlikely that they
would approach the Einstein ones asymptotically. In that case, we
could compute their conserved charges-defined as integrals at
infinity, using the definitions for generic quadratic models
\cite{dt1}. Let us stick to the quadratic models of the
form\footnote{Note that we changed normalization of the
cosmological constant compared to the previous section.}
 \be
  I = \int d^D\, x \sqrt{-g} \Big \{ { R\over 2
\kappa} + 2\Lambda_0 +\a R^2 + \b R^2_{\mu\nu} + \gamma (
R_{\mu\nu\rho\sigma}^2 -4R^2_{\mu\nu}+R^2 ) \Big \}\;.
\label{action}
 \ee
  This model allows constant curvature
spacetimes with an effective cosmological constant given as
 \be
\Lambda = -{1\over 4 f(\alpha,\beta, \gamma_) \kappa} \left \{ 1
\pm \sqrt{ 1 + 8 \kappa f(\alpha,\beta, \gamma) \Lambda_0} \right
\} \hskip 0.5 cm \mbox{for} \hskip 0.5 cm f(\alpha, \beta, \gamma
) \ne 0\;, \label{condition}
 \ee
 where
  \be
  f(\alpha, \beta, \gamma )
= { (D-4)\over (D-2)^2}(D\a +\b) + {\gamma (D-4)(D-3)\over
(D-2)(D-1) } \; .
 \ee
When the bare cosmological constant vanishes ($\Lambda_0 = 0$ ),
(A)dS spaces are still allowed and one has the $+$ sign branch in
(\ref{condition}).  Conserved charges in this model, which we
quote below, were defined in \cite{dt1}
 \be
 Q^\mu(\bar{\xi}) &=& \Big \{ { 1 \over \kappa} + {4 \Lambda D
 \alpha \over D-2} +{4 \Lambda
 \beta \over D-1  } + {4 \Lambda \gamma (D-4)(D-3)\over (D-2)(D-1)} \Big \}
 \int d^{D-1}\,x \sqrt{ -\bar{g}}
\bar{\xi_\nu}{\cal{G}}_L^{\mu\nu} \nonumber \\
&&+ (2\alpha +\beta) \int  dS_i \sqrt{-g} \Big \{ \bar{\xi}^\mu
\bar{\nabla}^i R_L + R_L \bar{\nabla}^\mu\, \bar{\xi}^i
- \bar{\xi}^i \bar{\nabla}^\mu R_L \Big \} \nonumber \\
&& +\beta \int dS_i \sqrt{-g}\Big \{ \bar{\xi}_\nu
\bar{\nabla}^{i} {\cal{G}}_L^{\mu \nu} - \bar{\xi}_\nu
\bar{\nabla}^{\mu} {\cal{G}}_L^{i \nu} - {\cal{G}}_L^{\mu\nu}
\bar{\nabla}^{i} \bar{\xi}_\nu + {\cal{G}}_L^{i \nu}
\bar{\nabla}^{\mu}\bar{\xi}_\nu \Big \}\;. \label{fullcharge}
 \ee
where ${\cal{G}}_L^{\mu\nu}$ and $R_L$ are the linear parts of the
Einstein tensor and the scalar curvature, respectively. The second
and the third line vanish for Einstein spaces. The first line, on
the other hand is just a factor times the cosmological Einstein
theory's charges (\ref{ad}). Therefore for asymptotic  Kerr-AdS
solutions, their conserved charges are given by the first term in
(\ref{fullcharge}), under the condition (\ref{condition}). Let us
specifically  consider the popular Einstein--Gauss--Bonnet theory,
$\alpha = \beta = 0$. Also, implementing the condition
(\ref{condition}) (with the + sign) we have,
 \be
 Q^\mu = - \sqrt{1 + 8\kappa
f(\gamma,0,0)\Lambda_0}{1\over \kappa}\int d^{D-1}\,x \sqrt{
-\bar{g}}\bar{\xi_\nu}{\cal{G}}_L^{\mu\nu} \; .
  \ee
Although the energy seems to have the wrong sign, this is a red
herring: As shown in \cite{db}, for the non-rotating case the
exact metric reads
 \be
 ds^2 = g_{00}dt^2 + g_{rr}dr^2 + r^2 d\Omega_{D-2}
  \ee
   \be
-g_{00} = g^{-1}_{rr} = 1+ {r^2\over 4\kappa
\gamma(D\!-\!3)(D\!-\!4)} \left \{ 1 \pm \left \{ 1 + 32\gamma
\kappa (D\!-\!3)(D\!-\!4){m\over (D\!-\!2) r^{D\!-\!1}}\right
\}^{1\over2}\right \} ,
 \ee
 whose asymptotic
forms branch into Schwarzschild and Schwarschild-de-Sitter
respectively, \be -g_{00} = 1- {r_0\over r^{D-3}}, \hskip 1 cm
-g_{00} = 1 +{4m\over(D-2)} r^{D-3} +{r^2\over \gamma(D-3)(D-4)}.
\ee We see that SdS branch comes with the ``wrong" sign compared to
the usual Schwarzschild one. Therefore, the minus sign in the energy
becomes positive once ${\cal{G}}_L^{\mu\nu}$ is explicitly computed.
We conclude that the  conserved charges in the
Einstein--Gauss--Bonnet theory for such asymptotic solutions would
be simply proportional to those of (\ref{finalen}, \ref{ang})
cosmological gravity:
 \be
 E_{\mbox{GB}}= \sqrt{1 + 8\kappa
f(\gamma,0,0)\Lambda_0}\,E_D \hskip 2 cm  J_i(\mbox{GB}) = \sqrt{1 +
8\kappa f(\gamma,0,0)\Lambda_0}\,J_i,
 \ee
It is important to note that if the coefficient $\sqrt{1 + 8\kappa
f(\gamma,0,0)\Lambda_0}$ does not vanish, then one can simply
rescale the Killing charges to get the Einstein charges
(\ref{finalen}, \ref{ang}).

\section{Conclusions}

Using the charge definitions via background Killing charges of
\cite{ad,dt1} we have computed the mass and angular momenta of the
rotating Kerr-AdS black holes for $D$ dimensions for cosmological
Einstein gravity.  As a test of stability, we checked that the
corresponding charge definitions for higher order would lead to the
same values for asymptotically similar geometries up to the
indicated constant rescaling.

\section{Acknowledgments}
The work of S.D.\ is supported by NSF grant PHY 04-01667; that of
B.T.\ by the ``Young Investigator Fellowship" of Turkish Academy of
Sciences (TUBA) and by a TUBITAK Kariyer Grant. B.T. thanks {\"
O}zg{\" u}r Sar{\i}o\u{g}lu for a fruitful discussion on TMG.

\myend